\documentstyle[aps,multicol,fleqn,psfig]{revtex}

\newcommand{\beq}{\begin{equation}}
\newcommand{\eeq}{\end{equation}}
\newcommand{\beqa}{\begin{eqnarray}}
\newcommand{\eeqa}{\end{eqnarray}}
\newcommand{\ket} [1] {\vert #1 \rangle}
\newcommand{\bra} [1] {\langle #1 \vert}

\newcommand{\mod}{~{\rm mod}~}

\begin{document}

\title{Greenberger-Horne-Zeilinger paradoxes for many qudits}

\author{Nicolas Cerf$^{1,2,}$\footnote[2]{ncerf@ulb.ac.be}, Serge
Massar$^{3,}$\footnote[1]{smassar@ulb.ac.be} and Stefano
Pironio$^{3,}$\footnote[5]{spironio@ulb.ac.be}}

\address{$^1$ Ecole Polytechnique, CP 165, 
Universit\'e Libre de Bruxelles, 1050 Brussels, Belgium\\
$^2$ Jet Propulsion Laboratory, California Institute of Technology, 
Pasadena, California 91109\\
$^3$ Service de Physique Th\'eorique, CP 225, 
Universit\'e Libre de Bruxelles, 1050 Brussels, Belgium}

\date{July 2001}
\draft
\maketitle

\begin{abstract}
We construct GHZ contradictions for three or more parties
sharing an entangled state, the dimension $d$ of each subsystem being an even
integer greater than 2. The simplest example that goes beyond the standard
GHZ paradox (three qubits) involves five ququats ($d=4$). 
We then examine the criteria a GHZ paradox must satisfy 
in order to be genuinely $M$-partite and $d$-dimensional.
\end{abstract}

\pacs{PACS numbers: 03.65.Bz, 03.67.-a, 89.70.+c}

\begin{multicols}{2}
\narrowtext

The entanglement of bipartite quantum systems of dimension greater than two
as well as the entanglement of multipartite quantum systems 
are questions far from being completely understood today,
and they motivate much of the current work in quantum information theory. 
One of the most important insights into multipartite (actually tripartite)
entanglement is provided by the Greenberger-Horne-Zeilinger (GHZ) 
argument\cite{ghz}. In its formulation given by Mermin\cite{M},
the GHZ argument is both an intrinsic contradiction arising when dealing with 
non-contextual variables (a Kochen-Specker theorem) and a Bell-EPR theorem 
that rules out local hidden-variable models. Furthermore, the GHZ argument
is an important primitive for building quantum information-theoretic 
protocols that decrease the communication complexity\cite{CB}, and it plays
a central role in the understanding of entanglement since the  
GHZ state is the maximally entangled state of three qubits\cite{G}.

In the present paper, we show how to construct GHZ contradictions for
three or more systems of dimension $d$ greater than 2 (qudits). 
In particular, we define several families of GHZ contradictions involving  
$M$ qudits that are based on operator relations, similarly to the
standard GHZ paradox. We also give precise conditions
that every GHZ paradox must fulfill in order to be genuinely $M$-partite 
and $d$-dimensional. This is of interest for the classification
of entanglement of multipartite and multidimensional systems: since GHZ
paradoxes provide an all-or-nothing refutation of local realism by quantum
mechanics, one expect that GHZ states are in some sense maximally entangled
states. Several extensions on the original work by GHZ and Mermin
have been proposed previously, as for example GHZ contradictions 
involving more than three qubits\cite{PRC}. More recently, it
also has been shown how to carry out a set of measurements on a multipartite
multidimensional system in a generalized GHZ state such that the 
correlation functions between the
measurement outcomes exhibit a contradiction with local hidden
variable theories of the GHZ type\cite{ZK}. 
However, in contrast to the present work, the
results of \cite{ZK} are not based on relations between a set of operators. 
Instead, our work more closely parallels Mermin's formulation of the GHZ
argument, being based on an algebra of operators. In particular, this implies
that each GHZ paradox presented in this paper is associated with a
(state-independent) KS theorem as well as a basis of GHZ states.
Our work is also related to multi-dimensional quantum error correcting
codes (the connection between quantum codes and GHZ contradictions 
has already been displayed for qubits in \cite{DiV}).

Let us consider a $d$-dimensional Hilbert space in which we define the 
{\em unitary} operators
\beqa
X&=&\sum_{k=0}^{d-1} \ket{(k+1) \mod d}\bra{k} \label{X} \\
Y&=&e^{i \pi p /d}\sum_{k=0}^{d-1} e^{2\pi i k/d} \ket{(k-1) \mod
  d}\bra{k} 
\label{Y}\\
Z&=&\sum_{k=0}^{d-1} e^{2\pi i k/d} \ket{k}\bra{k} \label{Z}
\eeqa
which satisfy $XY=e^{i \pi p /d}Z$, where $p=0$ for $d$ odd and $p=1$ for
$d$ even. These operators are (up to a phase) the error operators
that are used in multi-dimensional quantum error correcting codes \cite{Getal}.
For qubits ($d=2$), they correspond to the Pauli matrices: $X=\sigma_x$,
$Y=\sigma_y$, and $Z=\sigma_z$. 
The overall phases in Eqs. (\ref{X})-(\ref{Z}) are chosen so that these
error operators satisfy 
\beq\label{dpower}
X^d=Z^d= Y^d=\openone \ .\eeq
These operators also obey the commutation relations
\beq
Y^b X^a = e^{2\pi i ab/ d} X^a Y^b, \ 
Z^a X^b = e^{2\pi i ab/ d} X^a Z^b
\label{comm}
\eeq
for all integers $a,b$.

A simple example of a GHZ contradiction based on the
above operators consists of 5 parties each having a ququat 
(a 4-dimensional system). Consider the following 6 product operators:
\beqa \label{GHZ1}
V_0 &=& X_1 \otimes  X_2 \otimes   X_3\otimes   X_4\otimes   X_5\nonumber\\
V_1 &=& (X_1)^3 \otimes  Y_2\otimes   Y_3\otimes   Y_4\otimes   Y_5\nonumber\\
V_2 &=& Y_1 \otimes   (X_2)^3\otimes   Y_3\otimes   Y_4\otimes   Y_5\nonumber\\
V_3 &=& Y_1 \otimes   Y_2\otimes   (X_3)^3\otimes   Y_4\otimes   Y_5\nonumber\\
V_4 &=& Y_1 \otimes   Y_2\otimes   Y_3\otimes   (X_4)^3\otimes   Y_5\nonumber\\
V_5 &=& Y_1 \otimes   Y_2\otimes   Y_3\otimes   Y_4\otimes   (X_5)^3 \
\eeqa
One easily checks that these operators $V_i$ commute since $YX=i XY$,
so that they can all be simultaneously
diagonalized.  The eigenvalues of each $V_i$ are the 4th roots of
the identity since $V_i^4 = \openone$.
Furthermore, the product $V_0 V_1 V_2 V_3 V_4 V_5 = - \openone$, which
implies that the product of the eigenvalues of the 6 operators $V_i$
must be equal to $-1$. 
For instance a common eigenstate of the above operators 
with eigenvalues $V_0 = +1$, $V_1=V_2=\ldots =V_5 = -1$ 
is the generalized GHZ
state $|\Psi\rangle = {1 \over \sqrt{4}} 
\sum_{k=0}^3 |k\rangle\otimes |k\rangle\otimes
|k\rangle\otimes |k\rangle\otimes |k\rangle$.

Before presenting the KS and Bell-EPR forms of the GHZ argument associated with
these operators, let us note that we can always
associate an observable to a unitary operator 
$U=\sum u_i \ket{u}_i\bra{u_i}$, where $u_i$ and
$\ket{u_i}$ are the eigenvalues and eigenvectors of $U$. Indeed,
there is a one to one correspondence between $U$ and the Hermitian operator 
$H=i \log U=i \sum (\log u_i\ \mbox{mod}2 \pi) \ket{u_i}\bra{u_i}$.
By measuring $H$ and exponentiating the result, one can associate to $U$
a c-number (of unit norm) which will be one of its eigenvalue. We
shall call this the result of the measurement of $U$ in the following. 
Note that this remark does not apply for qubits as Pauli matrices 
are both Hermitian and unitary.

Let us now turn to the KS form of the GHZ contradiction (\ref{GHZ1}).
Suppose one tries 
to ascribe a definite value $v(V_k)$ to each of the operators
$V_k$. These operators are constrained by the relation $V_0 V_1 V_2 V_3 V_4 V_5
= - \openone$. Since they commute, the same relation must hold 
for their values:
\begin{equation}
v(V_0) v(V_1) v(V_2) v(V_3) v(V_4) v(V_5) = -1
\label{KS}
\end{equation}
Invoking non-contextuality, we can 
assign to the operator $V_k$ the
product of the values of the 5 one-party operators that appear in the tensor
product defining it. For instance, we have
\begin{equation}
v(V_1)=v(X_1)^3 v(Y_2)v(Y_3)v(Y_4)v(Y_5)
\end{equation}
Inserting this in (\ref{KS}) gives
\begin{equation}
v(X_1)^4v(Y_1)^4 \ldots v(X_5)^4v
( Y _ 5 ) ^ 4=-1
\label{KS2}
\end{equation}
Now the value associated to an operator must be one of its eigenvalues.
Equation (\ref{dpower}) therefore
implies that each of the $v(X)$ or $v(Y)$ must be a $4^{th}$
root of unity. Therefore the product on the left hand side of
Eq. (\ref{KS2}) is +1, although the right hand side is -1, 
so that the assignment of values is
impossible. This the content of the KS theorem.

The Bell-EPR form of the GHZ contradiction proceeds along the same line as the
KS form but with the noteworthy difference that the assignment of values to
each operators $X_j$, $Y_j$ is now justified by the weaker assumption of local
realism. Indeed, suppose the five parties separated from each other
and share a quantum state in a simultaneous eigenstate of the 5 operators
$V_0,\ldots,V_5$. For definiteness we take the state to be $|Psi\rangle$,
as defined above.
In principle one can learn
the result of the measurement of $X_j$ or
$Y_j$ by party $j$ by adequate measurements on the other four parties since
the product of the results must be one of the eigenvalues of $\Psi$:
$V_0=1$, $V_1=V_2=V_3=V_4 =V_5=-1$. 
Therefore, according to the EPR criterion of local realism,
one must assign to each
party $j$ a value $v(X_j)$ and $v(Y_j)$ for both the operators
$X_j$ and $Y_j$, which is one of the 4-th roots of the identity. Reasoning
as above, one then gets the same contradiction. In this way the GHZ argument
provides a very simple way to rule out local realism.

Let us now generalize the above GHZ contradiction to any odd number $M$ 
($\ge 3$) of parties, 
each having a qudit of dimension $d=M-1$. The corresponding GHZ
operators can be written as
\beq\label{example2}
\underbrace{
\left.
\begin{array}{ccccc}
X & X & X &  \ldots & X \\
X^{d-1} & Y & Y & \ldots  & Y \\
Y & X^{d-1} & Y & \ldots & Y \\
Y & Y & X^{d-1} & \ldots & Y \\
\vdots & \vdots & \vdots & \ddots & \vdots \\
Y & Y & Y & \ldots & X^{d-1}
\end{array}
\right\}
}_{\mbox{$M=d+1$ parties}}
\begin{array}{c}
\mbox{$M+1$}\\
\mbox{$=d+2$}\\
\mbox{ operators}
\end{array}
\eeq
where the columns correspond to the $M$ different parties
and the lines to the $M+1$ different operators.
We note that the generalized GHZ state $|\Psi\rangle$
is once more a common eigenstate of the $M+1$ operators,
giving rise to the same kind of contradiction.
This example can be further generalized by considering the $M+1$ operators
$W_0, W_1, \ldots, W_M$:
\beqa
W_0 &=& \underbrace{X^a \ldots  X^a}_{\mbox{$M$
    terms}}\nonumber\\
W_1 &=&  \underbrace{X^b \ldots  X^b}_{\mbox{$n$
    terms}}
\underbrace{Y^c \ldots  Y^c}_{\mbox{$p$
    terms}}
\underbrace{\openone \ldots  \openone}_{\mbox{$q$
    terms}}
\underbrace{Y^c \ldots Y^c}_{\mbox{$p$
    terms}}\label{WWWW}\\
W_k &=& \mbox{cyclic permutations of $W_1$} \ (1<k\leq M) \nonumber
\eeqa
where
\beq
2p=M-n-q
\label{2p}
\eeq
($M-n-q$ is thus even).
In order to have a GHZ paradox we require that:
\begin{list}{--}
\item the operators $W_j$ commute,
\item
\item if one assigns a classical value to the operators $X_j$
and $Y_j$ ($j=1,\ldots,M$), then the  product $v(W_0)\ldots v(W_M) =
+1$,
\item the product of operators $W_0W_1\ldots W_M \neq + \openone$.
\end{list}
\medskip

The first condition is already satisfied for
$j=1,\ldots M$ because of the cyclic permutations in the construction. The
requirement that $W_0$ also commutes with the other $W_j$'s imposes the
additional constraint $(e^{2\pi i ac/d})^{2p}=1$, or
\beq
2 p\, a\, c = k\, d
\label{cond1}
\eeq
where $k>0$ is an arbitrary integer. The second condition is satisfied if, in
each column, the number of $X$'s and $Y$'s is a multiple of $d$. This implies
that
\beq
2 p\, c=k'\, d
\label{cond2}
\eeq
and
\beq
n\, b + a=k''\, d
\label{cond2b}
\eeq
with $k',k''>0$ being arbitrary integers. [Note that Eq. (\ref{cond2}) implies
Eq. (\ref{cond1})]. The product of the M+1 operators $W_j$ is
\beq
W_0W_1\ldots W_M=e^{2\pi i [bcnp(M-n+1)/d]} \openone
\eeq
so that, using (\ref{cond2}), the third condition yields
\beq
b\, k'\, n\, (M-n+1)=2l+1
\eeq
where $l>0$ is an arbitrary integer. Thus $b,k',n$ and $(M-n+1)$ must be odd
integers. This implies that the number of parties $M$ must be odd, and, 
given Eq. (\ref{cond2}), that
the dimension $d$ must be even regardless of $c$. From Eq. (\ref{cond2b}), we
also have that $a$ must be odd, while Eq. (\ref{2p}) implies that $q$ is even.

As an illustration,
let us consider the special case $c=1$, $q=0$ and $k=1$.
Thus, for any even dimension $d$ and any odd $n$, there is a GHZ contradiction
for $M= d + n$ parties, with the exponent $a$ and $b$ given 
by Eq. (\ref{cond2b}). The operators given in
Eq. (\ref{example2}) are just the subclass $a=1$, $b=d-1$,
$n=1$. Another example is that of five qubits ($d=2$, $M=5$, $n=3$):
\beq \label{example3}
\underbrace{
\left.
\begin{array}{cccccc}
\ X\ &\ X\ &\ X\ &\  X\ &\  X\ \\
X & X & X &  Y &  Y \\
Y & X & X &  X &  Y \\
Y & Y & X &  X &  X \\
X & Y & Y &  X &  X \\
X & X & Y &  Y &  X
\end{array}
\right\}
}_{\mbox{$5$ parties}}
\begin{array}{c}
\mbox{6}\\
\mbox{ operators}
\end{array}
\eeq
The GHZ state here $|\Psi\rangle={1\over \sqrt{2}}
\ket{00000}+\ket{11111}$ is a common eigenstate of these operators
and gives rise to a paradox.

Other families of GHZ contradictions are also possible. For instance,
replacing $n=3$ and $q=0$ in the above example by $n=1$ and $q=2$ yields
\beq
\label{example4} \underbrace{
\left.
\begin{array}{cccccc}
\ X\ &\ X\ &\ X\ &\  X\ &\  X\ \\
X & Y & \openone & \openone & Y \\
Y & X & Y & \openone & \openone \\
\openone & Y & X &  Y & \openone \\
\openone &\openone & Y & X &  Y \\
Y &\openone &\openone & Y & X &
\end{array}
\right\}
}_{\mbox{$5$ parties}}
\begin{array}{c}
\mbox{$6$}\\
\mbox{ operators}
\end{array}
\eeq
which is the paradox obtained from the five-qubit error correcting 
code\cite{DiV}. Here, the logical states $\ket{0_L}$ and $\ket{1_L}$ of the
five-qubit code are GHZ states associated with this paradox.

Although $M$ was restricted to odd numbers in what precedes,
it is also possible to build GHZ contradictions with an even number
of parties. In \cite{PRC}, an example of qubits shared between 4 parties was
given. This example can be generalized to an even number $M$ of qudits of
dimension $d=M-2$ as follows:
\beq \label{example5}
\underbrace{
\left.
\begin{array}{ccccc}
X & Y^{d-1} & Y^{d-1} &  \ldots & Y^{d-1} \\
X^{d-1} & Y & Y & \ldots  & Y \\
Y & X^{d-1} & Y & \ldots & Y \\
Y & Y & X^{d-1} & \ldots & Y \\
\vdots & \vdots & \vdots & \ddots & \vdots \\
Y & Y & Y & \ldots & X^{d-1}\\
Y^{d-1} & X & X &  \ldots & X \\
\end{array}
\right\}
}_{\mbox{$M=d+2$ parties}}
\begin{array}{c}
\mbox{$M+2$}\\
\mbox{ operators}
\end{array}
\eeq
A common eigenstate of these operators is the GHZ state $|\Psi\rangle ={1 \over
\sqrt{d}}\sum_{k=0}^{d-1} e^{-i \pi k(k+2) /d} \ket{k}\otimes \ldots \otimes
\ket{k}$.

The above examples thus illustrate that it is possible to construct several
families of GHZ contradictions involving many parties, each sharing 
a high-dimensional system. We now examine with care what should be the
precise meaning of a {\em multipartite} and {\em  multidimensional} 
GHZ paradox. 


{\em Multipartite GHZ paradox:
A GHZ paradox  is genuinely $M$-partite if one cannot reduce the
number of parties and still have a paradox.}

This is best illustrated by an example. 
In \cite{PRC}, a GHZ paradox with 5 qubits was defined by the following
operators:
\beq
\underbrace{
\left.
\begin{array}{cccccc}
\ X\ &\ X\ &\ X\ &\  X\ &\  X\ \\
X & Y & Y &  X &   X \\
Y & X & Y &  Y &   Y \\
Y & Y & X &  Y &   Y 
\end{array}
\right\}
}_{\mbox{$5$ parties}}
\begin{array}{c}
\mbox{$4$}\\
\mbox{ operators}
\end{array}
\eeq
This paradox is not genuinely 5-partite according to our criterion. 
Indeed, these operators, restricted to
the first 3 parties, constitute a GHZ contradiction (in fact this is the
original paradox as formulated by Mermin).
Moreover, these operators, 
restricted to the last 2 parties, simply commute. As a consequence, the
eigenstates of these 4 operators can be written as tensor products of
states belonging to the first three parties times states belonging to
the last two parties. For instance the state ${|000\rangle +
|111\rangle \over \sqrt{2} } \otimes {|00\rangle + |11\rangle \over
\sqrt{2}}$ is a common eigenstate of these 4 operators. As a consequence, 
this state does not exhibit 5-partite entanglement.

{\em Multidimensional GHZ paradox:
a GHZ paradox is genuinely $d$-dimensional if one cannot reduce the
dimensionality of the Hilbert space of each of the parties to less
than $d$ and still have a paradox.}

More precisely consider a GHZ paradox defined by the $M$-partite
operators $W_k$ [e.g. those introduced in Eq. (\ref{WWWW})]. 
Suppose that there exist $M$ projectors $\Pi_l$ of rank less than $d$, 
each acting on the space of the $l$th party,
such that the operators
$\tilde W_k = \Pi_1 \otimes \ldots \Pi_M W_k \Pi_1 \otimes \ldots
\Pi_M$ define a lower-dimensional GHZ paradox. 
Then, the original paradox defined by 
these operators $W_k$ is not genuinely $d$-dimensional.
Let us illustrate this by a GHZ paradox in which 3 parties 
have a ququat (4-dimensional system), defined by the operators:
\beq\label{example6}
\underbrace{
\left.
\begin{array}{ccc}
X & X & X \\
X^3 & Y^2 & Y^2 \\
Y^2 & X^3 & Y^2 \\
Y^2 & Y^2 & X^3
\end{array}
\right\}
}_{\mbox{$3$ parties}}
\begin{array}{c}
\mbox{$4$}\\
\mbox{ operators}
\end{array}
\eeq
On the basis of the commutation relations (\ref{comm}) one could
expect that this is a genuinely 4-dimensional contradiction. Indeed,
the relation $YX= XY e^{i 2 \pi /d}$ can only be realized in a
Hilbert space whose dimension is at least $d$. [To prove this
suppose $X$ is diagonal: $X |k\rangle = e^{i\phi} |k\rangle$. Then the
commutation relation implies that the states $Y^p|k\rangle$
are also eigenstates
of $X$ with eigenvalue $e^{\phi+i 2 \pi p /d}$. 
Taking $p=1,\ldots,d$ yields $d$ distinct eigenvalues].
However in the example
(\ref{example6}), the operator $Y$ only appears to the power 2. Hence the only
commutators that are relevant to the paradox are $XY^2 = - Y^2 X$ and $X^3 Y^2
= - Y^2 X^3$ which can be realized in a 2-dimensional space. Using the
representations (\ref{X}) and (\ref{Z}), one sees that if one projects each
party onto the subspace spanned by the two vectors $\ket{0}+\ket{2}$ and
$\ket{1}+\ket{3}$, one still has a paradox. 
Thus the paradox (\ref{example6}) is not genuinely 4-dimensional, 
but only 2-dimensional.

All the multipartite multidimensional
GHZ contradictions that are exhibited in this paper are  
constructed from tensor products of
operators $X$ and $Y$ raised to different powers (with commutation relation
$Y^aX^b = X^bY^a e^{i 2 \pi ab/d}$).
Such a  paradox is
genuinely $d$-dimensional if, in each column (i.e. for each party),
the algebra generated by $X$ and $Y$ raised to the powers which appear 
in that column can only be represented in a Hilbert space of
dimension at least $d$.
(This was not the case in the last example since the 
algebra of the operators $\{X,X^3,Y^2\}$ could be represented in a 2
dimensional space.)

The above criteria guaranteeing that a GHZ paradox is genuinely
multipartite and genuinely $d$-dimensional are satisfied by all the examples
given in this paper [Eqs. (\ref{example2}), (\ref{example3}),
(\ref{example4}), and (\ref{example5})]. 
These criteria can also be applied to the general case Eq. (\ref{WWWW}). 
One would
then obtain additional conditions on the parameters $a$, $b$, and $c$. 
For instance,  the operators that appear in each column of
Eq. (\ref{WWWW}) are $\{X^a, X^b, Y^c\}$. The algebra generated by these
operators will be realized in a space of dimension at least $d$, so that the
paradoxes will be genuinely $d$-dimensional if $c$ and $d$ 
are relatively prime (i.e. their greatest common divisor is one),
and if $a$ or $b$ is relatively prime with $d$. 
To ensure that the first condition is satisfied, we
can take $c=1$. This is not restrictive since, if $c$ and $d$ are relatively
prime, there is a unitary operation that map $\{X^a, X^b, Y^c\}$ to $\{X^{a'},
X^{b'}, Y\}$, so that the algebra generated by the new set of operators is
identical to the one generated by the original set. Let us now examine 
the conditions that are necessary for the paradoxes in Eq. (\ref{WWWW})
to be genuinely multipartite. Removing
any number of columns (i.e. any parties), there are always two line $W_k$ and
$W_l$ such that $W_kW_l=e^{2 \pi i bc/d}W_lW_k$. Since $e^{2 \pi i bc/d}\neq 1$
because $c$ and $d$ are relatively prime and $b=1,\ldots d-1$, the condition
that all the operators $W_j$ must commute is not satisfied, so that the
remaining parties do not make a paradox. The generalization (\ref{WWWW})
is thus genuinely multipartite provided it is already genuinely
$d$-dimensional.

In summary, we have shown how to generalize the GHZ argument to
higher dimensional systems and more than three parties. Our method
is connected to the techniques used to construct error-correcting
codes for higher dimensional systems.
Interestingly, in all the GHZ-type paradoxes we have constructed, the
dimension is even and is strictly less than the number of
parties. We do not know whether this is necessarily the case, or if it 
is due to the restricted set of constructions we have considered. 
We have also seen that all the paradoxes one could naively expect 
to be multipartite and multidimensional are not necessarily so. In some cases, 
it is possible to reexpress the paradox in a
lower dimensional space, or, in other cases, the GHZ state associated with the
paradox can be represented as a product of states belonging to different 
subsets of parties. Finally, we have discussed criteria 
that ensure that a GHZ paradox
is truly $M$-partite and $d$-dimensional.

\end{multicols}
\end{document}